\documentclass[fleqn,usenatbib]{article}
\usepackage[utf8]{inputenc}
\usepackage[square,sort,comma,numbers]{natbib}
\usepackage{macros}
\usepackage[textwidth=15cm]{geometry}
\usepackage{hyperref}
\usepackage{authblk}
\usepackage{listings}
\title{MG-MAMPOSSt, a code to test gravity at galaxy-cluster scales: a technical introduction}
\author[1]{Developed by L. Pizzuti}
\author[2]{I. D. Saltas} 
\author[3,4]{A. Biviano}
\author[5]{G. A. Mamon}
\author[6]{L. Amendola}

\affil[1]{\small Osservatorio Astronomico della Regione Autonoma Valle d'Aosta,  Loc. Lignan 39, I-11020, Nus, Italy}

\affil[2]{\small CEICO, Institute of Physics of the Czech Academy of Sciences, Na Slovance 2, 182 21 Praha 8, Czechia
}
\affil[3]{\small INAF-Osservatorio Astronomico di Trieste, via Tiepolo 11, 34143 Trieste, Italy}
\affil[4]{\small IFPU - Institute for Fundamental Physics of the Universe, via Beirut 2, 34014 Trieste, Italy}
\affil[5]{\small Institut d’Astrophysique de Paris (UMR 7095: CNRS \& Sorbonne Universit{\'e}), 98 bis Bd Arago, F-75014 Paris, France}
\affil[6]{\small Institute of Theoretical Physics, Philosophenweg 16, Heidelberg University, 69120, Heidelberg, Germany}
\date{}

\usepackage{bm}
\usepackage{graphicx}
\usepackage[T1]{fontenc}
\usepackage{amsmath}
\usepackage{mathtools}
\usepackage{xcolor} 
\definecolor{grey}{rgb}{0.4,0.5,0.6}
\definecolor{darkgreen}{rgb}{0.0,0.5,0.0}

\begin{document}
\textwidth=10cm
\maketitle
\begin{abstract}
    The \textsc{MG-MAMPOSSt} code is a license-free \textsc{Fortran95} code to perform tests of General Relativity (GR) through the analyses of kinematical data of galaxy clusters based on the Jeans' equation. The code is based on the \textsc{MAMPOSSt} method, and extends the original code through new parametrisations of the gravitational potential for general families of gravity theories beyond GR aimed to explain dark energy. \textsc{MG-MAMPOSSt} is further supplemented with a new capability to produce weak lensing forecasts for joint kinematic+lensing analysis.
    
    This document provides a technical description of the code's new features, functionality with respect to the original version, and instructions on its installation and use. Finally, we explain how the code could be further modified to include a wider family of gravity models and/or density profiles, that could allow its application in broader theoretical frameworks as well as other physical systems such as stellar clusters. A detailed forecast analysis for the  modified gravity models currently implemented in the code can be found in the paper of Pizzuti et al., 2021.

\end{abstract}

\section{Introduction}
In view of the growing interest amongst the communities of cosmology and astrophysics to test the nature of gravity and dark energy theories at large scales, this manual is intended to provide a guide for the functionality and the basic usage of \textsc{MG-MAMPOSSt}, a \textsc{Fortran95} code capable of performing tests of gravity models with the kinematics of member galaxies in clusters. The code is based upon
the original \textsc{MAMPOSSt} method developed by G. Mamon, A. Biviano and G. Boué (Ref. \cite{Mamon01}, hereafter MAM13) \footnote{A public version of \textsc{MAMPOSSt} by G. Mamon can be found at \href{https://gitlab.com/gmamon/MAMPOSSt}{https://gitlab.com/gmamon/MAMPOSSt}.}.But whereas the original \textsc{MAMPOSSt} code relies on the assumption of a standard Newtonian gravitational potential, \textsc{MG-MAMPOSSt} implements general 
and viable models of gravity beyond General Relativity (GR), with the aim of placing constraints on their theory space at galaxy-cluster scales, as well as of investigating the essential statistical degeneracy between model parameters. In addition, the code is capable of producing complementary weak-lensing forecasts towards joint kinematics+lensing analyses.\footnote{The original \textsc{MAMPOSSt} has been used in a joint kinematics+lensing analysis by Ref. \cite{Verdugo+16}.}

\textsc{MAMPOSSt} (Modelling Anisotropy and Mass Profile of Spherical Observed Systems) determines mass profiles of galaxy clusters (or in general, spherical systems in dynamical equilibrium) by analysing the internal kinematics of the cluster members.\footnote{MAMPOSSt has also been used for elliptical galaxies traced by globular clusters and  dwarf spheroidals traced by their stars \citep{Mamon+15}, and has been recently extended into \textsc{MAMPOSSt-PM} to handle proper motions in star clusters (Ref. \citealp{Mamon&Vitral22}, see Ref. \citealp{Vitral&Mamon21}).} Given an input of projected positions and line-of-sight (l.o.s) velocities of the member galaxies, and under the assumptions of spherical symmetry and dynamical relaxation, the code solves the Jeans equation to reconstruct the gravitational potential, the velocity anisotropy profile, and (optionally) the projected number density profile. 

\textsc{MG-MAMPOSSt} extends the method to gravity scenarios beyond GR, where the gravitational potential is explicitly modified by the presence of an additional scalar degree of freedom, resulting in effective mass profiles different from GR. The resulting effective mass profile can be confronted against real or synthetic data provided as input to the code. In addition, the code requires a parametric modelling for the profiles of velocity anisotropy, mass and number density, with several available choices which can be tuned as input. The code's main output is a tabulated likelihood as a function of all the free model parameters of gravity and other input physics.

We emphasise that the goal of this document is to deliver the necessary instructions for the operation of the new features in \textsc{MG-MAMPOSSt}. We will defer from discussing the details of the original \text{MAMPOSSt} method on which the code is built upon, and we refer to Ref. MAM13 for an exposition of \textsc{MAMPOSSt}. What is more, for a detailed discussion of the application of \textsc{MG-MAMPOSSt} in theories beyond GR with synthetic data we refer to Ref. \cite{Pizzuti2021}.

The current licence-free version has been developed by L. Pizzuti based on the original 	\textsc{Fortran77} code of A. Biviano. I.D. Saltas, L. Amendola contributed to the theoretical background and equations as implemented in the code, assisted with the implementation of the statistical analysis in the code and advised in various other parts. G. Mamon contributed to the development of the weak lensing forecasts and to the refining of the user interfaces.

This document is organized as follows: In Section \ref{sec:preliminaries} we explain the basics around running the code, and in Section \ref{sec:generalities} we provide a brief overview of the method on which \textsc{MG-MAMPOSSt} and the original \textsc{MAMPOSSt} codes are based. In Section \ref{sec:models} we describe the new modified gravity models implemented in the code. In Sections \ref{sec:input1} and \ref{sec:options} we list all the input parameters that can be defined in \textsc{MG-MAMPOSSt}. In Section \ref{sec:output} we briefly discuss the output files of the code. Finally, Section \ref{sec:test} provides an illustrative example of a complete run, and an outlook is provided in Section \ref{sec:conc}.

\section{Preliminaries} \label{sec:preliminaries}
Running the code requires a \textsc{Fortran95} compiler or higher. The code can be downloaded from the following link\\
\\
\href{https://github.com/Pizzuti92/MG-MAMPOSSt}{https://github.com/Pizzuti92/MG-MAMPOSSt}.\\
\\
The downloaded folder contains the main source code (\texttt{gomamposstoptS.f}) and all the support files.  The configuration and installation requires at least CMake 3.17.1. 
In the main \textsc{MG-MAMPOSSt} folder execute the following commands:\\
\\
\texttt{mkdir build}\\
\texttt{cd build/}\\
\texttt{cmake ..}\\
\texttt{cmake} -\texttt{-}\texttt{build .}\\
\texttt{sudo cmake} -\texttt{-}\texttt{install .}\\
\texttt{cd ..}\\
\\
To run and test the code, type:\\
\\
\texttt{gomamposst < gomamposst\textunderscore x.inp}\\
\\
which produces the main outputs, or\\
\\
\texttt{./scripts/script\textunderscore runmam.sh  -}\texttt{-options} 
\\
which also generates additional plots if the MCMC mode is selected (see below). 
Note that to run the above script, permissions should be changed to make it executable. Otherwise, one can simply use the  \texttt{sh} command.  Installing the code and running \texttt{./scripts/script\textunderscore runmam.sh} without changing the input parameters and the operating options should produce the last example provided in Section \ref{sec:test}. The additional options are \texttt{-d <directorypath>}, which select the directory where the file \texttt{gomamposst\textunderscore x.inp} (see below) is located - default is the current working directory - and \texttt{-t} which prints the execution time of the code.

Input data should be stored in a text file named \texttt{datphys.dat} in the folder \texttt{data}. The file is structured as a table where the number of rows coincides with the number of data points. The first column is the projected radius in units of $\text{kpc}$, the second and thirds columns represent the l.o.s. velocities and the associated errors in units of $\text{km/s}$.
Note that the first two lines are considered as comment lines when \textsc{MG-MAMPOSSt} reads the data.

Input/output file names and locations are indicated in \texttt{gomamposst\textunderscore x.inp}; in particular, input files correspond to the first two lines. The role of \texttt{pars\textunderscore test.txt} (second line of   \texttt{gomamposst\textunderscore x.inp}) will be described in detail within Section \ref{sec:input1}, while the output files are summarized in Section \ref{sec:output}.

\section{Generalities}\label{sec:generalities}

\textsc{MG-MAMPOSSt} operates relying on the input of the projected phase space $(R,v_\text{z})$ of the cluster member galaxies. Here, $R$ is the projected distance from the cluster center at which a galaxy is seen by the observer, and  $v_\text{z}$ the velocity measured along the l.o.s. in the rest frame of the cluster. Only those tracers lying in the projected range $[R_\text{low},R_\text{up}]$ will be considered in the fitting procedure of the model parameters (see Section \ref{sec:input1}). As mentioned before, the current version of the code assumes parametric expressions for the various kinematical quantities. Moreover, the 3-dimensional velocity distribution is taken to be a Gaussian. The latter assumption has been well-tested through cosmological simulations, as explained in the original \textsc{MAMPOSSt} paper (MAM13). 

The output likelihood is computed for a given set of models and parameters from the solution of the spherical Jeans' equation (see e.g. Ref. \cite{MamLok05}),
\begin{equation}
\label{eq:sigmajeans}
\sigma^2_r(r)=\frac{1}{\nu(r)}\int_r^{\infty}{\exp\left[2\int_r^s{\frac{\beta(t)}{t}\text{d}t}\right]\nu(s)\frac{\text{d}\Phi}{\text{d}s}\text{d}s},
\end{equation}
projected in the phase space. The above equation captures the main necessary input required for the \textsc{MAMPOSSt} method. In particular, the gradient of the gravitational potential, which in turn contains information about the model of gravity, as well as the velocity anisotropy profile ($\beta(r)$, eq.~[\ref{eq:beta}]) and the projected number density of tracers ($\nu(r)$). The parametrisations for all these quantities implemented within the code are defined in the following sections.

\begin{figure}
    \centering
    \includegraphics[width=10cm]{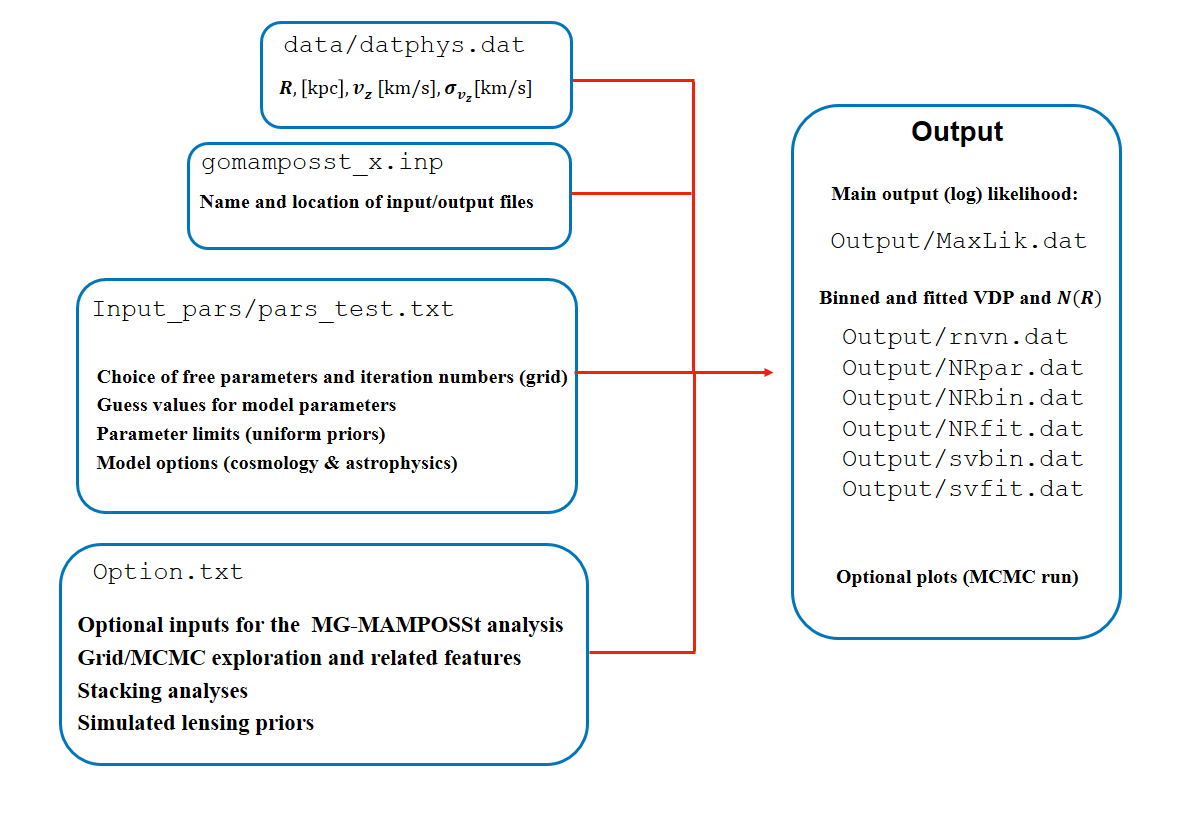}
    \caption{ Sketch of the \textsc{MG-MAMPOSSt} algorithm overviewing the hierarchy and connection between the input/output files in the code. Detailed information about each file is described in the text.}
    \label{fig:scheme}
\end{figure}

The current version of the code can handle up to a seven-dimensional parameter space, with two parameters defining the mass profile, one parameter for the velocity anisotropy profile, one for the number density profile, and finally, two parameters related to the modified gravity framework. Each parameter can be either treated as free in the fitting procedure, or it can be assigned pre-defined values (see also Section \ref{sec:input1}). The list of possible parameters and the relevant equations for the modified gravity parametrizations implemented are summarized in Table \ref{tab:params}.
\textsc{MG-MAMPOSSt} additionally returns a group of data files, stored in the \texttt{Output} folder, which contain the binned and fitted projected distribution of galaxies ($N(R)$), and the binned/fitted velocity dispersion profiles (VDPs), as explained in Section \ref{sec:output}. In Figure \ref{fig:scheme} we present a scheme summarizing all the relevant aspects in order to perform a complete run of the code.

A general analytic expression of the likelihood function can be found in MAM13.

\section{Implemented modelling} \label{sec:models}

\subsection{Mass and velocity anisotropy profiles}
In the original code of MAM13 there are several possible choices for the modelling of the dark matter mass profile in the $\Lambda$CDM scenario. \textsc{MG-MAMPOSSt} adds new parametrizations to handle popular modifications of gravity. At the moment, all the implemented non-standard profiles rely on the Navarro-Frenk-White (NFW, Ref. \cite{navarro97}) mass density profile to model the matter density distribution, which has been shown to provide a good description for simulated and observed galaxy clusters, both in GR and in modified gravity (see e.g. Refs. \cite{Umetsu_2020,Peirani17,Wilcox:2016guw}). Nevertheless, other mass models are going to be included in the code with upcoming versions. Since the mass profile, or equivalently, the gravitational potentials, enters only in the expression of the radial velocity dispersion eq. \eqref{eq:sigmajeans}, the implementation of new models can be performed directly by the user, modifying the subroutines where the above equation is involved in the auxiliary file \texttt{MAM.f}. In particular, the functions \texttt{sr2int(alr)} and \texttt{fa(tlog)} are the only parts of the \textsc{MG-MAMPOSSt} code where parametrizations for the mass profile appear. 

In the current version, ten choices of mass profiles are implemented: seven (GR) standard models, and three modified NFW models for MG frameworks. Each option in the code is identified by a specific acronym
as described also in Section \ref{sec:input1}. The implemented profiles are the following: NFW, Hernquist (Ref. \cite{Hernquist01}), Pseudo Isothermal Elliptical Mass Distribution (PIEMD, Ref. \cite{Kassiola93}), Burkert (Ref. \cite{Burkert01}), Softened Isothermal Sphere, Einasto with exponent $m=5$ (Ref. \cite{Einasto65}), "Linear Horndeski" NFW, Vainsthein screening (VS) NFW, Chameleon Screening (CS) NFW. We have further implemented a generalized NFW profile (gNFW hereafter) in GR, where the slope of the density is controlled by an exponent $\gamma$, equal to 1 for the standard NFW of eq. \eqref{eq:NFWdens}, which is now treated as a free parameter to be fitted together with the mass profile parameters. In this last case, the exponent $\gamma$ is given by the parameter $\mathcal{A}_1$. Further details are given in Section \ref{sec:input1} where all options in the code are over viewed.  

Except for the modified gravity profiles and the gNFW, each model is characterized by two parameters, namely a virial radius $r_{200}$, and a scale radius $r_\text{s}$. Though we do not give a detailed description of those profiles here, in the next subsection we will nevertheless present the key expressions for the modified gravity models implemented and the necessary parameters that need to be defined in the code.

In galaxy-cluster kinematics analyses, one of the major unknowns of the mass profile reconstruction is the so-called velocity anisotropy profile
\begin{equation}
    \beta(r)=1-\frac{\sigma^2_\theta(r)+\sigma^2_\phi(r)}{2\sigma^2(r)},
    \label{eq:beta}
\end{equation}
which accounts for the difference between the components of the galaxies' velocity dispersion. As for the mass profile, in the current version of \textsc{MG-MAMPOSSt} the velocity anisotropy is modeled as a parametric profile to be fitted within the procedure. Eight models have been implemented so far, identified by a specific label as for the mass modeling. Below we list the expressions of those models:
\begin{itemize}
     \item[1]  Constant anisotropy: $\beta(r)=\beta_C$,
    \item[2]  Mamon \& Lokas profile (Ref. \cite{MamLok05}):
    \begin{equation} \label{eq:ML}
        \beta_\text{ML}(r)=\frac{1}{2}\,\frac{r}{r+r_\beta}.
    \end{equation}
    \item[3] Osipkov-Merritt model (Refs. \cite{Osipkov79,Merritt85}):
    \begin{equation}
        \beta_\text{OM}(r)=\frac{r^2}{r^2+r^2_\beta}
    \end{equation}
    \item[4] Simplified Wojtak:
    \begin{equation}
     \beta_\text{SW}(r)=\frac{r^{3/2}}{r^{3/2}+r_\beta^{3/2}}.
    \end{equation}
    \item[5] Tiret model (Ref. \cite{Tiret01}):
    \begin{equation} \label{eq:tiret}
    \beta_\text{T}(r)=\beta_\infty\,\frac{r}{r+r_\beta}.
    \end{equation}  
    \item[6]  Modified Tiret ("Opposite") model (see e.g. Ref. \cite{Biviano01}):
    \begin{equation}
        \beta_\text{O}(r)=\beta_\infty\,\frac{r-r_\beta}{r+r_\beta}.
    \end{equation}
    \item[7] Generalized Osipkov-Merritt model (Refs. \cite{Osipkov79,Merritt85}, see also Ref. \cite{Mamon19}):
    \begin{equation}
        \beta_\text{gOM}(r)=\beta_0+(\beta_\infty-\beta_0)\frac{r^2}{r^2+r^2_\beta}
    \end{equation}
    \item[8] Generalized Tiret model (Ref. \cite{Tiret01}, see also Ref. \cite{Mamon19}):
    \begin{equation}
        \beta_\text{gT}(r)=\beta_0+(\beta_\infty-\beta_0)\frac{r}{r+r_\beta}
    \end{equation}
\end{itemize}

Note that models from one to four are characterized by only one free parameter, while models five and six have two parameters: $\beta_\infty$, the velocity anisotropy for $r\to\infty$, and $r_\beta$, the characteristic scale radius (anisotropy radius). Finally, the last two models requires an additional third normalization parameter $\beta_0$, which is the value of anisotropy at $r=0$ However, in \textsc{MG-MAMPOSSt} the anisotropy radius for models five to eight is assumed to be proportional to the scale radius of the mass profile $r_\beta\equiv \alpha\,r_\text{-2}$, where $\alpha=0.5,1.51,1$ for Hernquist model, Burkert model, and all the other models, respectively. This leaves a maximum of two free parameters for the orbit anisotropy to be fitted. Note that in the case of constant anisotropy, Tiret (normal, modified or generalized) and generalized Osipkov-Merritt, the normalization parameters used within the code are $\mathcal{A}_C=(1-\beta_C)^{-1/2}$,
$\mathcal{A}_\infty=(1-\beta_\infty)^{-1/2}$ and $\mathcal{A}_0=(1-\beta_0)^{-1/2}$ 
instead of $\beta_C$, $\beta_\infty$ and $\beta_0$ , respectively.

\subsection{Number density profiles}
The computation of the \text{MG-MAMPOSSt} likelihood requires the knowledge of the number density distribution of the tracers. Generally, this can be reconstructed directly from the phase space before running the code, but one can choose to fit the profile together with the anisotropy and the gravitational potential. Currently, \text{MG-MAMPOSSt} implements three models for $\nu(r)$, namely NFW model, eq. \eqref{eq:NFWdens}, Hernquist model and beta-model (Ref. \cite{Cavaliere78}), which can be selected from the input file \texttt{pars\textunderscore test.txt}
as explained in Section \ref{sec:input1}. All the models are specified by a normalization factor and a scale radius $r_\nu$. Moreover, the beta-model requires a (fixed) negative exponent $n_b$ which is set from input. Note that in eq. \eqref{eq:sigmajeans}, the normalization factor of $\nu(r)$ cancels out. Thus, the only relevant parameter of the profile is the scale radius $r_\nu$.

\subsection{NFW models for Vainsthein and Chameleon gravity}

The code is equipped with the two most popular and observationally viable  classes of dark energy models beyond GR based on a single, extra scalar field. These correspond to the so--called chameleon models (``{\bf Models I}") and Beyond Horndeski/DHOST models (``{\bf Models II}"). Common ground between the two families of models is the presence of the extra dynamical scalar degree of freedom ($\phi$) which introduces a new gravitational force. However, the structure of the fifth force  in each of them exhibits different characteristics. For a detailed exposition of these models and the associated equations we refer to our main paper (Ref. \citep{Pizzuti2021}), as well as to the original papers where the models were  first introduced (Refs.\citep{Kobayashi:2014ida,Crisostomi:2017lbg,Dima:2017pwp}).
Note that there is a third possibility implemented in the current version of the code: the "Linear Horndenski" models (``{\bf Models 0}"), characterized by a Yukawa-type fifth force without the inclusion of any screening mechanism. The model, first introduced in the code by Ref. \cite{Pizzuti:2017diz}, is actually a sub-case of the chameleon class, specified by two free parameters, namely the characteristic mass of the field $\mathcal{A}_1=m$ and the strength of the coupling $\mathcal{A}_2=\mathcal{Q}$. The difference between Models I and this linear framework is in the full treatment of the Yukawa suppression, which is neglected in the chameleon screening approximation. Models 0 will be removed in the upcoming versions of the code as its phenomenology at cluster scales is already encompassed by Models I.   

As shown in Ref. \cite{Pizzuti2021}, internal kinematics alone is generally not enough to provide stringent bounds on the modified gravity parameters, due to the strong degeneracy between model parameters. For this reason, \textsc{MG-MAMPOSSt} gives the possibility to include a simulated lensing information to your kinematics analysis in modified gravity, a feature which is particularly useful for forecasting the constraining power of the method. The structure of the lensing Likelihood depends on the chosen model of gravity and the details are given below.\\

{\bf Models I:} This is the family of chameleon models; they can be selected from \texttt{pars\textunderscore test.txt} as \texttt{M(r)=mNFW\textunderscore GC}. Their key characteristic is the existence of a scale $S$, the so--called screening radius, below (beyond) which the force is suppressed (unsupressed). The Newtonian potential gradients in the Jeans equation assumes the following form:
\begin{equation} \label{eq:pot_chameleon}
\frac
{\text{d}\Phi}{\text{d}r}=\frac{GM(r)}{r^2}+\frac{\mathcal{Q}}{M_\text{P}}\frac{\text{d}\phi}{\text{d}r},
\end{equation}
with $M(r)$ the total mass enclosed within a sphere of radius $r$, $M_\text{P}=(8\pi G)^{-1/2}$ is the reduced Planck mass, $G$ is the gravitational constant and $\mathcal{Q}$ a dimensionless coupling defining the strength of the chameleon force.  
For the scalar field $\phi$, we implement an analytic, approximated solution of eq. \eqref{eq:pot_chameleon}, which is explained in e.g. Ref. \cite{Terukina:2013eqa}, assuming the NFW model for the matter distribution. The NFW profile is defined through
\begin{equation}\label{eq:NFWdens}
\rho(r)=\frac{\rho_\text{s}}{r/r_\text{s}(1+r/r_\text{s})^2},
\end{equation}  
where the characteristic density $\rho_\text{s}$ is a function of $r_\text{s}$ and $r_{200}$. With this ansatz the field profile acquires the form
\begin{equation}
\label{eq:field}
\phi(x) =
  \begin{cases}
       \phi_\text{s}[x(1+x)^2]^{1/(n+1)}\simeq 0 &   r \le S  \\
     -\displaystyle{\frac{\mathcal{Q}\rho_\text{s}r_\text{s}^2}{M_\text{P}}}\,\displaystyle{\frac{\ln(1+x)}{x}}-\displaystyle{\frac{C}{x}}+\phi_{\infty} & r > S. 
  \end{cases}\
\end{equation}
In the last expression, $x \equiv r/r_\text{s}$, $\phi_{\infty}$ is the background value of the field and $C$ is an integration constant. Finally, $\phi_\text{s}<<\phi_\infty$ is the field value in the core of the matter distribution. Under the approximation that the interior solution is everywhere close to zero,  $C$ is fully determined by the free parameters of this gravity models (i.e. the fifth-force coupling $\mathcal{Q}$ and the cosmological value $\phi_{\infty}$)  and the mass profile parameters: $C = C(\mathcal{Q}, \rho_{s}, r_{s}, \phi_{\infty})$. The \emph{screening radius} $S$ represents the transition scale between the regime where the force is screened (i.e suppressed), and the one where the force is unscreened (i.e fully operating). The screening radius has a similar functional dependence, $S = S(\mathcal{Q}, \rho_{\rm{s}}, r_{\rm{s}}, \phi_{\infty})$. We refer to Ref. \cite{Pizzuti2021} for the explicit expressions. Note that, since the Yukawa suppression is neglected in this approximation, in order to avoid an overestimation of the fifth force the additional contribution to the gravitational potential is set to zero if $S>1.5\,R_\text{up}$, where $R_\text{up}$ is the maximum radius up to which galaxies are considered in the projected phase space of the \textsc{MG-MAMPOSSt} fit, as explained in the next sections. 

If the number of iteration for the coupling constant $n_{A2}$ (see Section \ref{sec:input1}) is set to  -1 for this class of models, the {\bf chameleon $f(\mathcal{R})$ sub-case} is selected, which correspond to a fixed $\mathcal{Q}=\sqrt{1/6}$. In this framework, the only free parameter is the so called scalaron field $f_\mathcal{R}=\partial f/\partial  \mathcal{R}$, where $\mathcal{R}$ is the Ricci scalar. The scalaron is connected to the chameleon field through:
\begin{equation}
\phi_\infty=-\sqrt{\frac{3}{2}}\ln(1+f_{\mathcal{R}0})M_\text{P}c^2_l\,,
\end{equation}
where $c_l$ is the speed of light. In \textsc{MG-MAMPOSSt} the first modified gravity parameter, generally dubbed as $\mathcal{A}_1$, corresponds to the background field  over the Planck mass $\phi_\infty/M_\text{P}$, given in units of $10^{-5}\, c_l^2$,    while the second modified gravity parameter is $\mathcal{A}_2=\mathcal{Q}$ (see \ref{sec:input1}).

 As for the lensing simulation, in chameleon gravity the fifth force doesn't affect photon propagation, except for a negligible rescaling of the gravitational constant $G$. Thus, lensing analyses are sensitive only to the GR parameters of the mass profile $r_\text{s},\,r_{200}$. As such, the additional information in \textsc{MG-MAMPOSSt} is provided as a bivariate Gaussian $P(r_\text{s},\,r_{200})$ whose parameters (namely, the means and the covariance matrix) are given as input by the user in the \texttt{Options.txt} file (see Section \ref{sec:options}).\\

{\bf Models II:} This class of models corresponds to the so--called Vainshtein breaking theories. The corresponding modified gravitational potential can be selected in \textsc{MG-MAMPOSSt} with \texttt{M(r)=mNFW\textunderscore BH}. Here, the fifth force within the spherical overdensity is only described by a single coupling parameter $\mathcal{A}_1=Y_1$. The Newtonian potential is defined through (e.g. \cite{Saltas:2019ius})
\begin{equation}
\frac{\text{d} \Phi}{\text{d}r} =  \frac{G M(r)}{r^2} +  \frac{Y_{1} G}{4} \, \frac{\text{d}^{2} M(r)}{\text{d}r^2}. \label{eq:BH1} 
\end{equation}
The dimensionless coupling $Y_1$ is the additional free parameter that can be fitted within the code. We note that, the second derivative $d^{2} M(r)/dr^2$ in equation (\ref{eq:BH1}) is expressed in terms of the density profile and its first derivative through $\text{d} M(r)/\text{d}r = 4 \pi \rho(r) r^2$. 

In these theories, the weak lensing potential, defined through 
$
\Phi_{\text{lens}} = \frac{1}{2}(\Phi + \Psi)
$,
is explicitly modified through the appearance of a new term in the relativistic potential $\Psi$ as
\begin{align}
\frac{\text{d} \Psi}{\text{d}r} =  \frac{G M(r)}{r^2} -  \frac{5\, Y_{2} G}{4\, r} \, \frac{\text{d} M}{\text{d}r}, \label{ippo-BH-Psi}
\end{align}
with $Y_2\equiv \mathcal{A}_2$ a dimensional coupling. Assuming the NFW mass model, the gradients of the potentials $\Phi,\,\Phi_\text{lens}$ read:
\begin{equation}  \label{BH-NFW}
\frac{\text{d}\Phi}{\text{d}r}=\frac{GM_\text{NFW}(r)}{r^2}\left\{1+\frac{Y_1}{4}\,\frac{r^2(r_\text{s}-r)/(r_\text{s}+r)^3}{[\ln(1+r/r_\text{s})-r/(r_\text{s}+r)]}\right\},
\end{equation}
\begin{equation} 
\frac{\text{d}\Phi_\text{lens}}{\text{d}r}=\frac{GM_\text{NFW}(r)}{r^2}\left\{1+\frac{\left[Y_1(r_\text{s}-r)-5\,Y_2(r_\text{s}+r)\right]}{8\,[\ln (1+r/r_\text{s})-r/(r_\text{s}+r)]}\frac{r^2}{(r_\text{s}+r)^{3}}\right\},
\end{equation}
where $M_{\text{NFW}}(r)$ is the standard NFW mass profile in GR, obtained by integrating eq. \eqref{eq:NFWdens}.

As for {\bf Models I}, in \textsc{MG-MAMPOSSt} we have implemented the possibility to forecast the constraints obtainable on a given MG model when adding a lensing information to the kinematics analysis only. 
In the case of Vainsthein screening, the code is able to perform a weak lensing simulation which produces a likelihood $\mathcal{L}_\text{lens}(r_\text{s},r_{200}, Y_1, Y_2)$ to be added to the kinematic likelihood. The simulation builds a (GR) mock reduced tangential shear profile assuming again a NFW model, and computes the log-likelihood as
\begin{equation}
    \ln\mathcal{L}_\text{lens}(\bm{\theta}_\text{l})=-\frac{1}{2}\sum_{i=1}^{N_\text{b}}\frac{\left[\langle g_\text{t}(R_i)\rangle-\langle g_{\text{t,vs}}(R_i|\bm{\theta}_\text{l})\rangle\right]^2}{\sigma^2_{\text{l},i}}.
\end{equation}
In the above equation, $\langle g_\text{t}(R_i)\rangle$ is the simulated averaged reduced tangential shear profile at projected position $R_i$, $\langle g_\text{t,vs}(R_i)|\theta_\text{l})\rangle$
is the theoretical profile computed for the Vainsthein model and the set of parameters $\bm{\theta}_\text{l}=(r_\text{s},r_{200}, Y_1, Y_2)$.
The number of bins is set to be $N_\text{b} = 10$, in agreement with current lensing surveys (e.g. Ref. \cite{Umetsu16}) in the projected radial range $(0.12\,R_\text{up},2.9\,R_\text{up})$, where $R_\text{up}$ is the maximum radius up to which tracers in the phase space are considered in the \text{MG-MAMPOSSt} fit (see Section \ref{sec:input1}). As for the uncertainties,
\begin{equation}\label{eq:errorlens}
\sigma^2_{\text{l},i}=\sigma^2_{\text{e},i}+\sigma^2_\text{lss},
\end{equation}
where $\sigma^2_{\text{e},i}=\sigma^2_\text{g}/[\pi(\alpha^2_\text{up}-\alpha^2_\text{low})n_\text{g}]$ is the noise due to the intrinsic ellipticity $\sigma^2_\text{g}$ of the sources lying within an annulus between the angles $\alpha_\text{low}$ and $\alpha_\text{up}$, and $\sigma^2_\text{lss}$ expresses the effect of the uncorrelated projected large scale structure. Finally, $n_\text{g}$ is is the average number of source galaxies per $\text{arcmin}^2$. The values of $\sigma_\text{g},\, \sigma_\text{lss}, \, n_\text{g}$ can be customized in \texttt{Options.txt}

\begin{table}
\centering
\begin{tabular}{ |c|c|c| }
 
 \hline
 {\bf Family of parameters} & {\bf Parameters} & {\bf Relevant equations}   \\
 \hline
Total matter profile (e.g. NFW)   &  $r_{200}$, $r_{\rm{s}}$   & (\ref{eq:NFWdens})  \\
Number density profile (e.g. NFW)   &  $r_{\nu}$  & (\ref{eq:NFWdens})  \\
Velocity anisotropy profile (e.g. Tiret) &  $\beta_{\infty}$ or $r_{\beta}$,  $\beta_{0}$  & (\ref{eq:tiret}) \\
Model 1 (Chameleon) & $Q$, $\phi_{\infty}$ & (\ref{eq:pot_chameleon}), (\ref{eq:field})    \\
Model 2 (Vainshtein)   & $Y_1$, $Y_2$ & (\ref{eq:BH1}), (\ref{ippo-BH-Psi})  \\
\hline
\end{tabular}

\caption{\label{tab:params}The main parameters associated with the modelling of the dark matter and velocity anisotropy profile, as well as the models of gravity implemented in \textsc{MG-MAMPOSSt}. They can be either pre-set in the code or left as free parameters to be fit when confronting against real or simulated data.}
\end{table}

\section{Overview of input parameters}
\label{sec:input1}
The file \texttt{input\textunderscore pars/ pars\textunderscore test.txt} contains the input parameters for the \textsc{MG-MAMPOSSt} procedure. All the parameters should be provided in a format \texttt{"<label> = <value>"}. The \texttt{"<label>"}s are {\bf mandatory} while the \texttt{"<value>"}s, if not given, are set by default, except for the case of the guess values of the model's free parameters. Inline comments can be inserted with a "\texttt{!}".
Input parameters are divided into four main groups: \\

{\bf 1. Number of iterations}, \\

{\bf 2. Guess values}, \\

{\bf 3. Model/setup options}, \\ 

{\bf 4. Parameter limits} .
\\ 

1. \textbf{Number of iterations} (integer): For the grid search mode, they represent the number of points in each free parameter over which the likelihood is computed. If set to $0, 1$, the parameter is fixed to its guess value, except for specific cases described below. The total number of likelihood evaluations is given by
\begin{equation}
    N_\text{tot}=\prod_i n_i, 
\end{equation}
where $i$ runs over the number of parameters which are fitted by the code. Currently, the maximum number of free parameters are seven.
\\ \\
When \textsc{MG-MAMPOSSt} is in MCMC mode, if \textbf{number of iterations} is different from zero, then the corresponding parameters are optimized within the chain. Otherwise the parameters are fixed to the guess value. 
\\ \\
In some specific cases other options are available; we will briefly describe them when listing all the input variables below.\\

2. \textbf{Parameter guess values} (real): They serve as an initial guess for the \textsc{MG-MAMPOSSt} fit both in grid and MCMC mode. If the corresponding number of iterations is set to zero, or one when in the grid-search option, the parameter guess is kept fixed within the code.\\

3. {\bf Model options} (real, integer, string): This family of values allow to select cosmological environment such as the value of the  Hubble parameter and the average redshift of the cluster, the mass or number density profiles, the velocity anisotropy profiles, as well as the optimization algorithm choices. \\

4. \textbf{Parameter limits} (real): They are labelled as $X^\text{low}$, $X^\text{up}$, where 
\begin{equation}
X\in\{r_{200},r_s,r_\nu,\beta,\mathcal{A}_1,\mathcal{A}_2\},
\end{equation} 
and are assumed as the limiting values of the parameter space. In the grid-search mode, they are considered only if the option \texttt{kpro} $=1$ (see \texttt{Options.txt} below). The grid exploration starts from the guess values and the likelihood is computed over a logarithmically-spaced set of points where the width is set such that the extremes corresponds to $X^\text{low}$ and $X^\text{up}$. If \texttt{kpro} $=0$, the width is fixed to a default value chosen optimally by the code for each parameter.
\\ \\ 
In the MCMC mode, the range within the limiting values should be assumed as a flat prior, i.e. 
\begin{equation}
   P(X)= \frac{1}{X^\text{up}-X^\text{low}} \begin{dcases}
        0 & X < X^\text{low}  \\
        1 & X^\text{low}\leq X\leq X^\text{up} \\
        0 & X > X^\text{up} .\\
    \end{dcases}
\end{equation}
The pre-factor $1/(X^\text{up}-X^\text{low})$ ensures the normalization of the probability distribution. One should be warned that, for Vainsthein screening, some values of $Y_1,\,Y_2$ can give rise to a negative effective cluster mass at some radii, which is physically inconsistent. Thus, in the MCMC run, in addition to the flat prior all the couples $(Y_1,\,Y_2)$ related to negative masses are discarded. 
If \texttt{kpro} $\ne 1$ the parameter limits are chosen by default.
\\

Below we list all the input parameters in the same order as they appear in \texttt{pars\textunderscore test.txt}: 

\begin{itemize}

    \item[-] \texttt{nr200} (integer): Number of iteration for the virial radius $r_{200}$.
    
    \item[-] \texttt{nrc} (integer): Number of iteration for the tracers scale radius $r_\nu$. For $\texttt{nrc}=-1$, it  assumes that $r_\nu$ is equal to the scale radius of the mass profile $r_s$ in the fit ({\bf option: ``Light Follows Mass"}). If $\texttt{nrc}=-2$ the code first fits the number density profile alone and the best fit found is then used as fixed value in the \textsc{MG-MAMPOSSt} procedure. 
    \item[-] \texttt{nrs} (integer): Number of iteration for the mass profile scale radius $r_s$.  For $\texttt{nrs}=-1$  assumes that $r_s$ is equal to the scale radius of the number density profile $r_\nu$ in the fit ({\bf option: ``Mass Follows Light"}). If $\texttt{nrs}=-2$ the mass scale radius is computed by using the theoretical relation of Ref. \cite{Maccio08} ($\Lambda$CDM option).
    \item[-] \texttt{nbeta} (integer): Number of iteration for the first anisotropy parameter. If $\texttt{nbeta}=-1$ the anisotropy profile is forced to be a Mamon \& Lokas profile (eq. \ref{eq:ML}) with $\beta\equiv r_\beta=r_s$. If $\texttt{nbeta}=-2$ the
    Hansen-Moore model of Ref. \cite{Hansen2006} is assumed ($\beta(r)$ related to $\rho_m(r)$).
    
    \item[-] \texttt{nA1} (integer): Number of iteration for the first MG parameter $\mathcal{A}_1$.
    
    \item[-] \texttt{nA2} (integer): Number of iteration for the second MG parameter. For the case of general chameleon gravity, $\mathcal{A}_2$ corresponds to the coupling constant $\mathcal{Q}$. If $n_{A2}=-1$ it forces the case of $f(\mathcal{R})$ gravity ($\mathcal{Q}=1/\sqrt{6}$).
    \item[-] \texttt{nbeta2} (integer): Number of iteration for the second anisotropy parameter, used only if \texttt{beta(r)}=\texttt{gT},\texttt{gOM}
    
    \item[-] \texttt{r200g} (real): guess starting value of the characteristic "virial" radius for the mass profile $r_{200}$, measured in units of $[\text{Mpc}]$. 

    \item[-] \texttt{rcg}(real): guess starting vaule for the scale radius of the number density profile  $r_\nu$,  units of $[\text{Mpc}]$. 
    
    \item[-]  \texttt{rsg}(real): guess starting value for the scale radius of the mass profile $r_s$, in  units of $[\text{Mpc}]$. 

    \item[-] \texttt{betag} (real): starting guess for the first velocity anisotropy profile parameter $\beta$ . If the selected model is "constant", "Tiret", "Modified Tiret", "generalized Tiret" or "Generalized Osipkov-Merritt" (see below) the parameter is dimensionless, and corresponds to $\mathcal{A}_{\infty/0}$. Otherwise it will be interpreted in units of $[\text{Mpc}]$.
    
    \item[-] \texttt{A1g} (real): Initial guess for the first modified gravity parameter $\mathcal{A}_1$ (see Section \ref{sec:models}). If the mass model is \texttt{mNFW\textunderscore GC}, the parameter is given in units of $10^{-5}c_l^2$, where $c_l$ is the speed of light (e.g. a value of $10$ means $10^{-4}c_l^2$). For the case of gNFW model, $\mathcal{A}_1=\gamma$ is the free exponent which characterizes the profile.
    
    \item[-]\texttt{A2g} (real): Initial guess for the second modified gravity parameter $\mathcal{A}_2$ (see Section \ref{sec:models}).
    
    \item[-] \texttt{beta2g} (real): starting guess for the second velocity anisotropy profile parameter $\beta_0$. Used only for the generalized Tiret and the generalized Osipkov-Merritt models. It is always dimensionless.
    
    \item[-] \texttt{H0} (real): The value of the Hubble parameter evaluated at redshift $z=0$, measured in units of $[ \text{km} \text{\, s}^{-1}\,\text{Mpc}^{-1}]$. Default is 70.
    
    \item[-] \texttt{za} (real): Average redshift of the cluster's center of mass. Default is 0.0.
    
    \item[-] \texttt{Olam} (real): Value of the $\Omega_\Lambda$ density parameter today. Default is 0.7.
    
    \item[-]\texttt{Omegam} (real): Value of the $\Omega_\text{m}$ density parameter today. Default is 0.3.
    
    \item[-] \texttt{Rlow} (real): Inner projected radius $R_\text{low}$, defining the minimum projected radius from the cluster center (given in $[\text{Mpc}]$) at which galaxies should be considered in the \textsc{MG-MAMPOSSt} fit. Default is 0.05.
    
    \item[-] \texttt{Rup} (real): Outer projected radius $R_\text{up}$, defining the maximum projected radius from the cluster center (given in $[\text{Mpc}]$) at which galaxies should be considered in the \textsc{MG-MAMPOSSt} fit. Default is the (mandatory) value  of \texttt{r200g}.
    
    
    \item[-] \texttt{N(R)} (string): Number density profile model. It selects the model for the tracer's projected number density profile. In the current version three possible choices are allowed, indicated by the following labels:\\
    \texttt{pNFW} for projected NFW,\\ \texttt{pHer} for projected Hernquist,\\ \texttt{beta} for beta-model.\\
    Default is \texttt{pNFW}.
    
    \item[-] \texttt{al} (real): Exponent for beta-model. The beta-model choice requires an additional input exponent which should be negative. The input parameter is ignored otherwise. Default is 0.0.
    
    \item[-] \texttt{M(r)} (string): Mass profile/gravitational potential model. It selects the allowed parametrisations of the mass profile to be used in the Jeans' equation. The labels corresponding to the implemented models are:\\
    \texttt{NFW} for NFW,\\ 
    \texttt{Her} for Hernquist,\\ 
    \texttt{PIEMD} for PIEMD,\\
    \texttt{Bur} for Burkert,\\
    \texttt{SoftIS} for Softened Isothermal Sphere,\\
    \texttt{Eis5} for Einansto with $m=5$,\\
    \texttt{gNFW} for generalized NFW,\\ 
    \texttt{mNFW\textunderscore LH} for modified NFW in Linear Horndeski gravity,\\
    \texttt{mNFW\textunderscore BH} for modified NFW in Vainsthein screening (Beyond Horndeski),\\
    \texttt{mNFW\textunderscore GC} for modified NFW in general Chameleon screening gravity.\\
    Details on each modified gravity model are given in Section \ref{sec:models}. Default is \texttt{NFW}.
    
    \item[-] \texttt{Beta(r)} (integer): Anisotropy profile model. It selects the  
    velocity anisotropy $\beta(r)$ in the Jeans' equation. The allowed profiles (see Section \ref{sec:models})  are: 
    \texttt{C} for constant anisotropy, \\  
    \texttt{ML} for Mamon \& Lokas, \\
    \texttt{OM} for Osipkov-Merritt, \\
    \texttt{WJ} for simplified Wojtak, \\
    \texttt{T} for Tiret, \\
    \texttt{O} for modified Tiret (Opposite), \\ 
    \texttt{gT} for generalized Tiret,\\
    \texttt{gOM} for generalized Osipkov-Merritt.\\
    Default is \texttt{C}.
    
    \item[-] \texttt{rcut} (real): Truncation radius. Value of the truncation radius needed in the pseudo-isothermal elliptical mass distribution (PIEMD). It is ignored for other mass profile models.
    
    \item[-] \texttt{FASTMODE} (integer): If equal to 1, the likelihood is estimated by using a grid of values (default 60 points) in the phase space $(R,v_z)$ of the galaxies and then bispline-interpolating over the data points. 
    
    \item[-] \texttt{OPT} (integer): \texttt{Optimization choice}. If required, an optimization algorithm is launched before the grid/MCMC likelihood computation to find the minimum of $-\ln \mathcal{L}$. Eventually, the resulting best fit parameters are used as guess values for the subsequent parameter space exploration. Currently, three choices are available:  \textsc{BOBYQA} ($\texttt{OPT}=0$), \textsc{NEWUOA} ($\texttt{OPT}=1$) or \textsc{POWELL} ($\texttt{OPT}=2$)\footnote{We point out that the POWELL algorithm does not work efficiently when considering the gT and gOM models for the velocity anisotropy, due to round-off errors in the integration routines, while the other two alghoritms do.}. For details, see MAM13. If $\texttt{OPT}=-1$ the optimization is skipped. 
    
    \item[-] \texttt{screen} (integer): \texttt{Screening mode} (available only for \texttt{M(r)=mNFW\textunderscore LH}). In linear Horndeski, one can choose to adopt the $f(\mathcal{R})$ sub-case, $\texttt{screen}=0$, where $\mathcal{A}_2$ is fixed to $\sqrt{1/6}$. In this framework, there is the possibility to include a model-dependent screened $f(\mathcal{R})$ model with Hu\&Sawicki functional form,implemented by assuming a simple analytical approximation. The transition between the screened and linear regime can be instantaneous ($\texttt{screen}=1$), or smoothed with an additional parameter controlling the sharpness ( $\texttt{screen}=2$). For $\texttt{screen}=-1$, the general linear Horndeski with two free parameters is selected.
    
    \item[-] \texttt{r2low}: Lower limit for $r_{200}$ $[\text{Mpc}]$. 
    \item[-] \texttt{r2up}: Upper limit for $r_{200}$ $[\text{Mpc}]$. 
    \item[-] \texttt{rclow}: Lower limit for $r_{\nu}$ $[\text{Mpc}]$.
    \item[-] \texttt{rcup}: Upper limit for $r_{\nu}$ $[\text{Mpc}]$.
    \item[-] \texttt{rslow} Lower limit for $r_\text{s}$ $[\text{Mpc}]$. 
    \item[-]\texttt{rsup}: Upper limit for $r_\text{s}$ $[\text{Mpc}]$.
    \item[-] \texttt{blow}: Lower limit for the first anisotropy parameter $\beta$. 
    \item[-] \texttt{bup}: Upper limit for the first anisotropy parameter $\beta$.
    \item[-]\texttt{A1low}: Lower limit for the first modified gravity parameter. 
    \item[-] \texttt{A1up}: Upper limit for the first modified gravity parameter.  
    \item[-] \texttt{A2low}: Lower limit for the second modified gravity parameter. 
    \item[-]\texttt{A2low}: Upper limit for the second modified gravity parameter.  
    \item[-] \texttt{b2low}: Lower limit for the second anisotropy parameter $\beta_0$. 
    \item[-] \texttt{b2up}: Upper limit for the second anisotropy parameter $\beta_0$.
    
\end{itemize}
Note that the parameter limits in the last block have as default values $10^-4$ and $10^2$ for all the lower bounds and upper bounds respectively.
\section{Operating Options} \label{sec:options}

The file \texttt{Options.txt} contains various options and switches for the new features in \textsc{MG-MAMPOSSt}. These are mostly related to the numerical analysis and evaluation of the posterior likelihood. Notice that, the input parameters can be \texttt{binary integers} (with values $0$ or $1$), \texttt{integers*4} or \texttt{reals*8}. 
As for the case of \texttt{pars\textunderscore test.txt}, all the parameters must be given in a format \texttt{"<label> = <value>"}. The \texttt{"<label>"}s are {\bf mandatory} while the \texttt{"<value>"}s, if not given, are set by default.  \\

\begin{itemize}
    \item \texttt{nmcmc (binary)}. Select between grid-search mode ($= 0$), and MCMC sampling ($ = 1$). The default value is $ = 1$.
    
    \item \texttt{Nsample (integer)}. Number of points in the MCMC run. Default is $400000$.
    \item \texttt{nlens (binary)}. Lensing information: If equal to 1, it adds to the kinematics likelihood, a probability distribution simulating additional information such as provided by a lensing mass profile reconstruction. For each set of values of the parameters, the joint (log) likelihood is then computed as
    \begin{equation}\label{eq:jointlike}
    \ln\mathcal{L}_\text{joint}=\ln\mathcal{L}_\text{dyn}+\ln\mathcal{L}_\text{lens},
    \end{equation}
    where $\ln\mathcal{L}_\text{dyn}$ and $\ln\mathcal{L}_\text{lens}$ are the log-likelihoods of the MG-MAMPOSSt procedure and the simulated lensing distribution respectively. 
    For linear Horndeski and Chameleon screening, where photon propagation is not affected by the new degrees of freedom, the lensing likelihood has the form of a bivariate Gaussian distribution $\mathcal{L}_\text{lens}=P_\text{lens}(r_\text{s},r_{200})$, specified by central values, standard deviations and correlation (see below). In Vainsthein screening, where lensing mass profile is explicitly modified by the MG parameters, the likelihood is computed by simulating a full tangential shear profile, as explained in Ref. \citep{Pizzuti2021}. The default value is $ = 0$.
    
    \item \texttt{r200t (real)}. "True" value of the cluster's virial radius (in unit of $\text{Mpc}$) around which the lensing distribution is centered. This are mandatory. In case it is not provided, the code switches to $\texttt{nlens}=0$.
    
    \item \texttt{rst (real)}. "True" value of the cluster's scale radius (in unit of $\text{Mpc}$) around which the lensing distribution is centered.
    This are mandatory. In case it is not provided, the code switches to $\texttt{nlens}=0$.
    
    \item \texttt{delta1 (real)}. For Vainshtein screening, it represents the intrinsic ellipticity of galaxies in the (weak) lensing simulations. For Chameleon screening, it is the relative uncertainty on the virial radius in the Gaussian distribution $\sigma_{r_{200}}/r_{200}$. Default for Vainsthein screening is $=0.3$, default for Chameleon screening is $=0.1$.
    
    \item \texttt{delta2 (real)}. For Vainshtein screening, it represents the large scale structure contribution to the errors in the (weak) lensing simulations. For Chameleon screening, it is the relative uncertainty on the scale radius in the Gaussian distribution $\sigma_{r_\text{s}}/r_\text{s}$. Default for Vainsthein screening is $=0.005$, default for Chameleon screening is $=0.3$.
    
    \item \texttt{delta3 (integer/real)}. For Vainshtein screening, it represents the number of galaxies per arcminute square in the (weak) lensing simulations. For Chameleon screening, it is the correlation $\rho$ in the Gaussian distribution. Default for Vainsthein screening is $=30$, default for Chameleon screening is $=0.5$.
    
    \item \texttt{kpro (binary)}. If it is equal to 1, the parameter space exploration is made over a given interval $[X^\text{low},X^\text{up}]$, where $X$ indicates a generic free parameter, as mentioned in Section \ref{sec:input1}. Default is 1. 

    \item \texttt{Nclust (integer)}. \textsc{MG-MAMPOSSt} allows for efficient statistical forecast analyses of the constraints on the implemented MG models. In particular, it is possible to input \texttt{Nclust} realizations of phase spaces at the same time to compute directly the joint likelihood for a given set of parameters, obtained from the combination of the likelihood from each single data-set.  These data files should be located in "\texttt{/data}" folder and named as \texttt{datphys\textunderscore<i>.dat}, where \texttt{<i>} labels the number of the file in ascending order, starting from 2 (e.g. \texttt{datphys\textunderscore2.dat}, \texttt{datphys\textunderscore3.dat}, ...). The file format is the same as the main input \texttt{datphys.dat} (see Section \ref{sec:generalities}). Default is $ = 1$. Note that in order to obtain meaningful results using this option, all the data files should be a realization of the same cluster (i.e. characterized by the same true values of all parameters).
    
    \item \texttt{nsingle (binary)}. If it is equal to 1, the \texttt{Nclust}-clusters likelihood is computed by simply multiplying by \texttt{Nclust} the likelihood from a single data-set. Useful to forecast a fast estimation of the limiting behaviour of the constraints when increasing the number of clusters. Default is $ = 0$.
    \end{itemize}
    
{\bf Additional Options} 
    \begin{itemize}
    
        \item \texttt{istop (binary)}. When equal to 1, the program stops after the preliminary optimization algorithm if the relative difference between the best fit found and the guess values is larger than $\vec{\epsilon}=(\epsilon_{r_{200}},\epsilon_{r_\nu},\epsilon_{r_\text{s}},\epsilon_{\beta},\epsilon_{\mathcal{A}_1},\epsilon_{\mathcal{A}_2},\epsilon_{\beta_0})$, or if the relative difference of the logarithm of the likelihood, computed at the best fit and at the guess values is larger than $\Delta_{\ln(\mathcal{L})}$ (see below). Default value is 0.
        \item \texttt{teps (array, real)}.
        Threshold values for $\vec{\epsilon}$. Default values are: 
    \begin{displaymath}
    \epsilon_i=0.1.
    \end{displaymath}
    \item \texttt{delik (real)}. Threshold value for the relative difference of the logarithmic likelihood \begin{displaymath}
    \Delta_{\ln(\mathcal{L})}=\frac{\ln(\mathcal{L})_\text{bf}-\ln(\mathcal{L})_\text{guess}}{\ln(\mathcal{L})_\text{guess}},
    \end{displaymath}
    where $\ln(\mathcal{L})_\text{bf}$ is the best fit likelihood according to \textsc{MG-MAMPOSSt} and $\ln(\mathcal{L})_\text{guess}$ is the likelihood computed at the parameters' guess values. Default is $= 0.2$.
    
        \item \texttt{nskip (binary)}. For the grid case, if equal to 1, the exploration starts from the free parameters guess values even if the  preliminary optimization is not skipped. This could be useful in modified gravity frameworks where the \textsc{MG-MAMPOSSt} likelihood presents more than one maximum, due to statistical degeneracy between model parameters. One can be interested in knowing the position of the global peak but sampling the likelihood only around a specific local maximum. Default is $=0$
        
        \item \texttt{nres (binary)}. For the grid exploration without fixed bounds (\texttt{kpro}$=0$), it selects a larger (0) or a smaller-size (1) grid step. Note that the steps are different for each parameter and adjusted depending on the number of available tracers in the fit, unless specified (see below). Default is $=0$
    
        \item \texttt{nequ (binary)}. For the grid exploration without fixed bounds (\texttt{kpro}$=0$), if equal to 1, it removes the re-scaling of the grid steps by the number of available tracers. Default is $=0$
    
    \item \texttt{nsame (binary)}. If equal to 1, likelihoods are computed on specific values of the parameters, given from an external input file. The name of the file should be \texttt{MaxLik}\textunderscore\texttt{input}.\texttt{dat}, structured as a table with seven columns (one for each parameter), following the same order as in the output likelihood \texttt{MaxLik}.\texttt{dat}, i.e. $r_{200},\, r_\nu,\, r_\text{s},\, \beta,\, \mathcal{A}_1,\, \mathcal{A}_2,\,\beta_{0}$ (see Section \ref{sec:output}).\\
    This option only works if the MCMC mode (\texttt{nmcmc}$=1$) is selected. Default is $=0$

\end{itemize}

\section{Output} \label{sec:output}
The \textsc{MG-MAMPOSSt} code produces several output files which are automatically saved in the \texttt{Output} folder. 
Below we list the files in the same order as they appear in the \texttt{gomamposst\textunderscore x.inp} script, summarizing their content and focusing in particular on the output likelihood/posterior\footnote{Note that we refer to ``likelihood" when no fixed bounds are assumed (\texttt{kpro}$=0$), i.e. no prior information, and to ``posterior" otherwise.}.
\begin{itemize}
    \item[] \texttt{rnvn.dat}. Projected distances $R_i$ (in $\text{Mpc}$), velocities $v_{z,i}$ (in $\text{km/s}$) of the tracers used in the fit and the associated probability $p(R_i,v_{z,i})$ to find a galaxy in the phase space $(R_i, v_{z,i})$ given the theoretical model. Note that the (log) likelihood is given by the sum of $\ln p$ over the tracers.
    
    \item[] \texttt{NRpar.dat}. Single-row data file where the choice of the number density profile, the best fit value of the scale radius $r_\nu$ and the value of the negative exponent for the beta-model ($k_\text{n}=3$) are stored. 
    
    \item[] \texttt{NRbin.dat}. Binned projected number density profile $N_\text{bin}(R)$ where the projected positions are in units of $\text{Mpc}$. The third column lists the associated errors. 
    \item[] \texttt{NRfit.dat}. Fitted projected number density profile $N_\text{fit}(R)$ where the projected positions are in units of $\text{Mpc}$.
    \item[] \texttt{MaxLik.dat}. File containing the tabulated likelihood/posterior computed by the \textsc{MG-MAMPOSSt} procedure. This is the main output of the code. The table is structured as follows: columns from 1 to 7 store the values of the parameters, ordered as $r_{200},\, r_\nu,$ $ r_\text{s},\, \beta,\, \mathcal{A}_1,\, \mathcal{A}_2,\, \beta_{0}$. Column 8 is $-\ln{\mathcal{L}}$, 
    while the integer number in the last column is related to the choice of the anisotropy profile, according to the following: 0='C', 1='ML', 2='OM', 21='gOM', 3='WJ', 4='T', 41='gT', 5='O'. The two rows at the end of the file highlight the minimum found by the exploration and the minimum found by the preliminary optimization respectively, which may not coincide for distributions with more than one local maximum.
    
    \item[] \texttt{svbin.dat}. Binned velocity dispersion profile (VDP). Each row shows the following quantities: number of galaxies in the i-th bin, $R_i$ (Mpc), VDP$(R_i)$ (km/s), VDP's upper error (km/s), VDP's lower error (km/s) 
    \item[] \texttt{svfit.dat}. Fitted VDP obtained by using the best fit values found by the \textsc{MG-MAMPOSSt} procedure. First column: projected position (in Mpc). Second column: VDP (km/s). 
\end{itemize}
In addition, running \texttt{./script\textunderscore runmam.sh} produces plots of the marginalized distributions for the free parameters {\bf if the MCMC exploration is selected}. Plots are generated with 
the \texttt{plot.py} Python3 script which makes use of the \textsc{GetDist} package
\footnote{\href{https://getdist.readthedocs.io/en/latest/index.html}{https://getdist.readthedocs.io/en/latest/index.html}}, a license-free Python package developed by A. Lewis (Ref. \cite{Lewis:2019xzd}). For some mass models, the script produces also a plot of the best fit mass profile with the associated one sigma and two sigma bands. The models for which this feature is implemented are NFW, Burkert, Hernquist, generalized NFW, and the modified gravity NFW for Vainsthein and Chameleon screening.

\section{A simple walk-through tutorial} 
\label{sec:test}
In this section we provide a tutorial on how to use the main functionality of the code to analyse kinematics and simulated weak-lensing data. 

In particular, we show an example of a complete run for the Vainsthein screening case (\texttt{mNFW\textunderscore BH}) using a single data-set shipped together with the code base. 
The data file is a mock phase space drawn from an isolated, spherically symmetric distribution of collisionless particles whose velocity dispersion is given by the solution of the Jeans' equation for a certain mass profile and velocity anisotropy model (see Refs. \cite{Pizzuti:2019wte,Pizzuti2021}). In particular, the data-set consists in $\sim 2700$ particles obtained from a NFW profile in GR (i.e. the fiducial value of the MG parameters is 0) with $r_{200}=1.4\, \text{Mpc}$, $r_\text{s}=0.33\, \text{Mpc}$ and $r_\nu\equiv r_\text{s}$. The phase space is obtained requiring that $\sim 600$ tracers are included in a sphere of radius $r_{200}$. As for the velocity anisotropy profile, we adopt the Tiret model of eq. \eqref{eq:tiret}, with $\beta_\infty=0.5$, $\beta_0=0.0$ and $r_\beta=r_\text{s}$.

Let us first consider the kinematical data only, without the additional lensing simulation. For comparison, we perform a grid and an MCMC parameter space exploration. In Table \ref{tab:times} we indicate the computational time of the two modes on a common laptop as a function of the number of free parameters.

As a first step, we open the \texttt{gomamposst\textunderscore x.inp} script  and adjust the name and the location of the data-set file. In our case we type:\\
\texttt{data/datphys\textunderscore test.dat}. Then, in \texttt{input\textunderscore pars/pars\textunderscore test.txt}, write all the correct settings for the case of interest. For simplicity, we assume as guess values the true values of the cluster's parameters. The {\bf second block of parameters} in the input file should appear as:\\
\begin{lstlisting}
************ free parameters initial guess values *********************

r200g = 1.41 ! mass profile r200 initial guess (Mpc)  
             
rcg = 0.33  ! N(R) scale radius initial guess (Mpc)  
             
rsg = 0.47  ! mass profile scale radius initial guess (Mpc)  
             
betag = 1.41 ! Anisotropy initial guess, 
             ! beta_C, a_ML, a_OM, a_W, beta_inf   
             
A1g = 0.00 ! first MG parameter initial guess     
             
A2g = 0.00  ! second MG parameter initial guess    
             
beta2g = 1.0 ! Second Anisotropy parameter initial guess beta0 for 
             ! gOM and gT models
************************************************************************
\end{lstlisting}

As for the cosmology, the synthetic halo is generated at average redshift 0 with $H=70\, \text{km/s\,Mpc}^{-1}$, $\Omega_\Lambda=0.7$ and $\Omega_m=0.3$. For the upper and lower limits of the fit, we consider $R_\text{up}=r_{200}$ and $R_\text{low}=0.05 \, \text{Mpc}$. For a detailed discussion about the physical meaning of this limits, see e.g. Refs. \cite{Mamon01, Biviano01, Pizzuti:2017diz}.

We consider the modified NFW model for Vainshtein screening  gravity of eq. \eqref{BH-NFW}, \texttt{mNFW\textunderscore BH}, with a Tiret anisotropy profile '\texttt{T}' and a NFW for the tracers number density distribution  \texttt{pNFW}. We adopt the fast-mode and the optimization algorithm NEWUOA, a derivative-free method developed by Ref. \citep{Powell06}, \texttt{OPT}=1,  
to find a preliminary best-fit. The remaining parameters are dummy in this case. Thus, the third block of parameters is given by 
\\
\begin{lstlisting}
************  Model options *******************************************

H0 = 70.0   ! Hubble constant at z=0   
                           
za = 0.0    ! average redshift of the cluster (needed to evaluate Hz)  
                                                     
Olam = 0.7   ! Omega lambda density parameter  
             
Omegam = 0.3  ! Omega matter density parameter 
             
Rlow = 0.05  ! Inner projected radius for sample selection (Mpc)  
             
Rup = 1.41   ! Outer projected radius for sample selection (Mpc)  
             

N(R) = pNFW
             ! model for number density profile N(R)
             ! Here we list the labels for N(R) and 
             ! then the corresponding profile:
             ! pNFW=projected NFW / pHer=projected Hernquist / 
             ! beta=beta-model
    
al = -0.00
             ! N(R) negative exponent (only if knfit=3)

M(r) =  mNFW_BH		  
             ! mass profile model. 
             ! Here we list the labels for M(r) and 
             ! then the corresponding profile:  
             ! NFW=NFW/ Her=Hernquist/ PIEMD=PIEMD/ Bur=Burkert/ 
             ! SoftIS=SoftIS/! Eis5=Einasto_m=5/ 
             ! mNFW_LH = mod_NFW linear Horndeski/
             ! mNFW_BH = mod_NFW beyond Horndeski/
             ! mNFW_GC= mod_NFW general chameleon/
             ! gNFW = generalized NFW
              
Beta(r) = T   
             ! Anisotropy model.
             ! Here we list the labels for beta(r) and
             ! then the corresponding profile: 
             ! C=Constant/  ML=Mamon&Lokas/ 
             ! OM=Osipkov-Merritt/  WJ=simplified Wojtak/
             ! T=simplified Tiret/ gT= generalized Tiret
             ! O= (Opposite) modified Tiret/ 
             ! gOM= generalized Osipkov-Merritt
              
rcut= 1.0     
             ! PIEMD model rcut in Mpc (only used if M(r)=PIEMD) 

FASTMODE= 1  
             ! Run MG-MAMPOSSt in fast mode interpolating over data points

OPT= 1        
             ! optimization algorithm: 0/1/2=bobyqa/newuao/powell
             ! -1 skip optimization
              
screen= 0   
             ! for linear f(R) kmp.eq.7, one can decide to set 
             ! an instantaneous transition between 
             ! screeening and linear regime, by using the 
             ! analytical approximation of Lombriser+12.                
             !-1/0/1/2=noscreen (general Hordenski)/noscreen f(R)/
             ! screen(instantaneous transition)
             ! /screen (arctan transition)         
			   
************************************************************************
\end{lstlisting}
Finally, one may select the parameter limits (priors) for the exploration in the case of \texttt{kpro}$=1$. For this example run we choose the intervals as (see Section \ref{sec:input1} for the ordering details):\\
\begin{lstlisting}
********* parameter limits *********************************************

r2low= 0.4   !r200 lower bound 
              
r2up= 5.0    !r200 upper bound 
              
rclow= 0.05  !rc lower bound 
              
rcup= 3.9    !rc upper bound
 
rslow= 0.04  !rs lower bound 
              
rsup= 3.9    !rs upper bound 
              
blow= 0.5    !beta lower bound
              
bup= 7.1     !beta upper bound 
              
A1low = -0.67 !first MG parameter lower bound 
              
A1up =   7.2  !first MG parameter upper bound
              
A2low=  -12.0 !second MG parameter lower bound
              
A2up= 12.0   !second MG parameter upper bound    
              
b2low = 0.5  !second beta parameter lower bound 
              
b2up = 7.1   !second beta parameter upper bound
              
************************************************************************
\end{lstlisting}
Before going to the \texttt{Options.txt} file, choose the free parameters in the first group of quantities  in\texttt{pars\textunderscore test.txt}. We recall that, for the MCMC mode, a value different from zero means that the corresponding parameter is optimized within the chain, while in the grid mode this value indicate the number of grid points where the likelihood is evaluated. Below the example setup for five free parameters $(r_{200},\,r_\text{s},\,\beta,\,\mathcal{A}_1,\,\mathcal{A}_2)$:\\
\begin{lstlisting}
**************  number of  steps in free parameters ********************
nr200 = 40  ! number of steps for r200 fit     
          
nrc = 0     ! number of steps for rc fit, scale radius of N(R)
                !   [if = 0 takes guess value]
                !   [if = -1 forces LfM, c_nu=c]
                !   [if = -2 fit N(R) outside MAMPOSSt]   
                
nrs =  40   ! number of steps for rs fit, scale radius of M(r)
                !   [if = 0 takes guess value]
                !   [if = -1 forces MfL, c=c_nu]
                !   [if = -2 forces LCDM, c=c(M)]       
          
nbeta = 40   ! number of steps for anisotropy parameter
                !   [if = -1 forces a_ML=r_s]
                !   [if = -2 forces Hansen+Moore]   
           
nA1 = 40     ! number of steps in the first MG parameter 
          
nA2 = 40     ! number of steps in the second MG parameter
             ! If equal to -1 force the case of chameleon f(R) gravity      
          
nbeta2 = 0  ! number of steps for the second anisotropy parameter                             
************************************************************************
\end{lstlisting}

In \texttt{Options.txt} we can now select the grid or MCMC search switch; let's first consider the grid exploration, \texttt{nmcmc} $=0$.
We furthermore set \texttt{kpro}$=1$ and \texttt{nlens}=0 (i.e. we do not include the lensing simulation for now). We do not modify the other parameters which are irrelevant in this first case. Important: we stress again that while the \texttt{<value>} can be omitted to ensure the default choice, all the \texttt{<labels>} described in Section \ref{sec:options} are mandatory, do not remove them from the file. 

The code can be now executed form a linux terminal by means of the command lines mentioned in Section \ref{sec:preliminaries}. Since the grid mode has been selected, no plots are shown at the end of the run. After $\sim 0.1$ seconds ($\sim 1$ second for normal mode)  from execution, the terminal should print the following lines:\\

\begin{lstlisting}
 optimiz. results: r200,rc,rs,beta,A1,A2,cbe0,-logL
 1.406  0.330  0.369  1.569  0.083  0.000  1.000   5908.378
 likelihood of the initial value and Delta chi square:
 r200,rc,rs,beta,A1,A2, -logL, Delta chi square:
 1.410  0.330  0.470  1.410  0.00   0.000  1.000   5909.734  2.711
\end{lstlisting}
Then, the parameter exploration starts around the minimum found and the values of the posterior/likelihood are written on \texttt{MaxLik.dat} for each set of parameters\footnote{In Vainsthein screening, kinematics is not sensitive to $Y_2$, which means that the likelihood does not depend on this parameter.}. Here is an example of the first lines of the tabulated likelihood for the case of five free parameters:\\
\\
      0.42844  \hspace{4pt}      0.33000  \hspace{4pt}     0.04000  \hspace{4pt}      0.50000    \hspace{4pt}   -0.67000    \hspace{4pt}  -12.00100 \hspace{4pt} 1.00000  \hspace{4pt}  9331.69590  \hspace{4pt}  4\\
      0.42844  \hspace{4pt}      0.33000  \hspace{4pt}      0.04580   \hspace{4pt}     0.50000    \hspace{4pt}   -0.67000   \hspace{4pt}   -12.00100  \hspace{4pt} 1.00000  \hspace{4pt}  9298.36094 \hspace{4pt}  4\\
      0.42844  \hspace{4pt}      0.33000    \hspace{4pt}    0.05245   \hspace{4pt}    0.50000   \hspace{4pt}    -0.67000    \hspace{4pt}  -12.00100 \hspace{4pt}  1.00000  \hspace{4pt}  9265.74595 \hspace{4pt}  4\\
      0.42844  \hspace{4pt}      0.33000   \hspace{4pt}     0.06005   \hspace{4pt}     0.50000   \hspace{4pt}    -0.67000  \hspace{4pt}    -12.00100  \hspace{4pt} 1.00000  \hspace{4pt}  9234.35828 \hspace{4pt}  4\\
      $\dots$\\
      \\
The exploration is made by means of nested cycles over the parameters of interest. The cycles  are ordered in the following way, from the innermost to the outermost: $r_\text{s}\to r_\nu \to \beta_0 \to \beta \to \mathcal{A}_1 \to \mathcal{A}_2 \to r_{200}$. In Table \ref{tab:times} we can see how the computational time becomes extremely large when increasing the number of free parameters, even when the fast mode is selected. Moreover, the structure of the table makes the output file enormous and very difficult to handle. Thus, the MCMC mode is suggested when the number of free parameters is $\ge 4$.  Note that in general the execution depends also on the number of data points where the likelihood is computed. However, since for real clusters the number of tracers can be around $\sim 10^3$ in very few optimistic cases, this factor has a mild influence on the total time.
\begin{table} 
\centering
\begin{tabular}{|c|c|c|c|c|c|}
   \hline
  \multicolumn{6}{|c|}{{\bf Computational time needed}}\\
  \hline
     &  \multicolumn{5}{c|}{Number of model parameters to fit} \\ 
     \hline
{\bf Mode} & 1 & 2 & 3 & 4 & 5 \\
\hline
\hline
grid  & 0.3 s  & 10 s  & 605 s  & $5\times10^4$ s  & $\sim 2$ weeks  \\
\hline
grid  fast & 0.3 s  & 2 s  & 106 s  & $6\times10^3$ s  & $\sim 2.5$ days  \\
\hline
MCMC  & $1.6 \times 10^4$ s  & $1.6 \times 10^4$ s  & $1.7 \times 10^4$ s  & $1.7 \times 10^4$ s  & $1.7 \times 10^4$ s  \\
\hline
MCMC fast & $1.3\times 10^3$ s  & $1.4 \times 10^3$ s  & $1.4 \times 10^3$ s  & $1.4 \times 10^3$ s  & $1.4 \times 10^3$ s\\
\hline
\end{tabular}
\caption{\label{tab:times} The required computational time in order to perform a complete run of \textsc{MG-MAMPOSSt} for the various modes as a function of the free parameters to be fitted. The grid search assumes 40 points for each parameter (i.e. $40^n$ total points with $n$ the number of free parameters), while the MCMC has been made sampling $10^5$  points every time. The run has been performed on an i7-8565U processor laptop (8th Gen) at $1.80$ GHz, with 16 GB RAM.}

\end{table}

To select the MCMC mode, set \texttt{nmcmc} $=1$ in \texttt{Options.txt}. As for the number of samples in the run, we choose $10^5$ points by typing
\texttt{Nsample} $=100000$ in the same file. Leave all the other options unchanged, save the file and run again the executable \texttt{./script\textunderscore runmam.sh}. Now the tabulated likelihood looks like as:
\\
\\
      1.43094   \hspace{4pt}     0.33000   \hspace{4pt}      0.32708  \hspace{4pt}       1.52905   \hspace{4pt}      0.07640  \hspace{4pt}      -0.04110  \hspace{4pt} 1.00000  \hspace{4pt}   5894.40454 \hspace{4pt}   4 \\
      1.38378   \hspace{4pt}      0.33000  \hspace{4pt}     0.31959  \hspace{4pt}    1.67060   \hspace{4pt}   0.24586 \hspace{4pt}   -0.03153  \hspace{4pt}   1.00000  \hspace{4pt}  5894.64370 \hspace{4pt}  4 \\
      1.34847 \hspace{4pt}   0.33000   \hspace{4pt}   0.32015 \hspace{4pt}     1.60624  \hspace{4pt}   0.14262 \hspace{4pt}   -0.06625 \hspace{4pt}    1.00000  \hspace{4pt}  5895.73375 \hspace{4pt}  4 \\
      $\dots$\\
      \\

To include the weak lensing simulation in Vainsthein screening, open again \texttt{Options.txt} and choose \texttt{nlens} $=1$, together with the related parameters. When \textsc{MG-MAMPOSSt} is run, for each vector of parameters the probability distribution is then computed with eq. \eqref{eq:jointlike}. In this example, we consider \texttt{r200t} $=1.4$ Mpc,  \texttt{rst} $=0.33$ Mpc, togehter with \texttt{delta1} $=0.25$, \texttt{delta2} $=0.005$ and \texttt{delta3} $=30$ for the intrinsic ellipticity, the large-scale structure noise and the number of galaxies per $\text{arcmin}^2$, respectively. An {\bf important note} is in order here: the weak lensing mock analysis is performed by assuming a value $z=0.44$ for the average cluster redshift. We further note that the background cosmology is almost irrelevant for our simulation, as in the kinematics analysis the phase spaces are always given in the cluster's rest frame.
\begin{figure}
    \centering
    \includegraphics[width=0.9\textwidth]{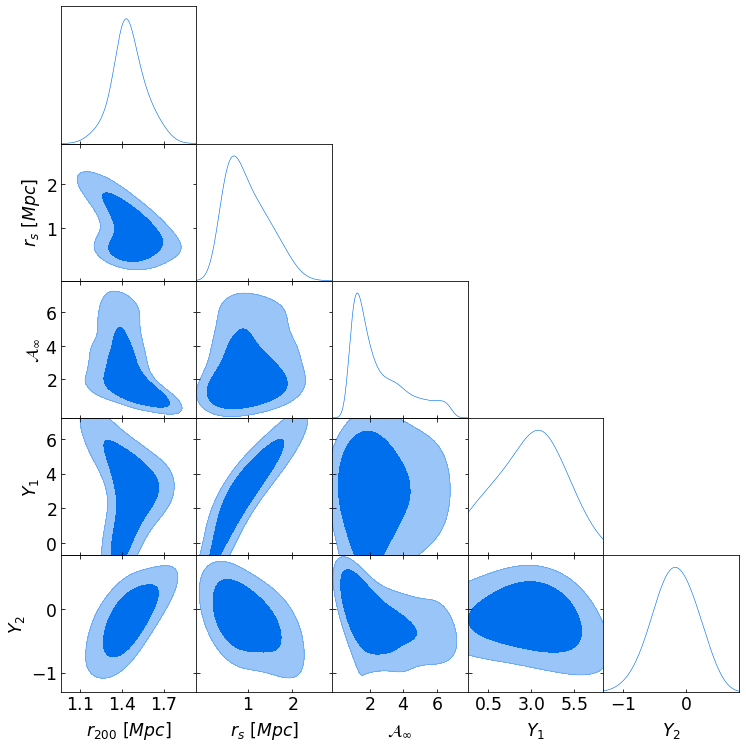}
    \caption{Marginalized distributions obtained from the output posterior of the \textsc{MG-MAMPOSSt} procedure in Vainsthein screening. The distributions refer to a run with five free parameters and kinematics+lensing analysis. The darker and lighter shaded areas represent the $1\sigma$ and $2\sigma$ regions, respectively}
    \label{fig:plots}
\end{figure}
In the case of an MCMC exploration, at the end of the \textsc{MG-MAMPOSSt} routine, the marginalized distributions are plotted, as shown in Figure \ref{fig:plots} for the case of a joint lensing+kinematic analysis in Vainsthein screening and with five free parameters.

\section{Summary}
\label{sec:conc}
We provided a detailed reference guide for the key features and usage of the \textsc{MG-MAMPOSSt} code. The code is based on the original \textsc{MAMPOSSt} method and code, and presents with important new features aiming at testing large classes of gravity models with the internal kinematics of galaxy clusters. What is more, \textsc{MG-MAMPOSSt} is now capable to complement kinematical analyses with additional, simulated lensing information towards consistency tests of gravity based on the reconstructed kinematical and lensing mass. In addition, the code is well-suited for analyses based on either synthetic or real data, and provides fast optimisation techniques for parameter exploration.

Although \textsc{MG-MAMPOSSt} is equipped with the currently most general theories for dark energy based on new scalar fields, there is certainly a big space for improvement. For example, it would be interesting to extend the theories currently implemented to new ones or add efficient parametrisations for the gravitational field, or other number/matter density profiles. Also, a very interesting, yet challenging task would be its extension for the case of other physical systems such as stellar clusters and dwarf galaxies. 
In this regard, we hope that the code will provide a useful tool for future analyses of gravity within the communities of cosmology and astrophysics, and see further extensions and applications. 

\section*{Acknowledgements}
LP is partially supported by a 2019 "Research and Education" grant from Fondazione CRT. The OAVdA is managed by the Fondazione Cle\'ment Fillietroz-ONLUS, which is supported by the Regional Government of the Aosta Valley, the Town Municipality of Nus and the "Unite\' des Communes valdotaines Mont-E\'milius.
I. D. Saltas is supported by the Grant Agency of the Czech Republic (GAČR), under the grant number 21-16583M. The authors further acknowledge S. Sartor for useful comments and suggestions, as well as all the developers of the free \textsc{Fortran} routines used in \textsc{MG-MAMPOSSt}. Credits are given in the header of each routine.

\bibliographystyle{ieeetr}
\bibliography{references}
\end{document}